\documentclass{article} % For LaTeX2e
\usepackage{iclr2021_workshop,times}
\usepackage[normalem]{ulem}
\usepackage{amsmath,amsfonts,bm}

\def\eqref#1{equation~\ref{#1}}
\def\1{\bm{1}}

\def\ra{{\textnormal{a}}}

\def\rx{{\textnormal{x}}}

\def\rva{{\mathbf{a}}}

\def\erva{{\textnormal{a}}}

\def\ervx{{\textnormal{x}}}

\def\rmA{{\mathbf{A}}}

\def\vmu{{\bm{\mu}}}
\def\vtheta{{\bm{\theta}}}
\def\va{{\bm{a}}}

\def\ve{{\bm{e}}}

\def\vx{{\bm{x}}}

\def\eva{{a}}

\def\mA{{\bm{A}}}

\def\mH{{\bm{H}}}
\def\mI{{\bm{I}}}
\def\mJ{{\bm{J}}}

\def\mX{{\bm{X}}}

\def\mSigma{{\bm{\Sigma}}}

\DeclareMathAlphabet{\mathsfit}{\encodingdefault}{\sfdefault}{m}{sl}
\SetMathAlphabet{\mathsfit}{bold}{\encodingdefault}{\sfdefault}{bx}{n}
\newcommand{\tens}[1]{\bm{\mathsfit{#1}}}
\def\tA{{\tens{A}}}

\def\tX{{\tens{X}}}

\def\gG{{\mathcal{G}}}

\def\sA{{\mathbb{A}}}
\def\sB{{\mathbb{B}}}

\def\sS{{\mathbb{S}}}

\def\emA{{A}}

\newcommand{\etens}[1]{\mathsfit{#1}}

\def\etA{{\etens{A}}}

\newcommand{\E}{\mathbb{E}}

\newcommand{\R}{\mathbb{R}}

\newcommand{\KL}{D_{\mathrm{KL}}}
\newcommand{\Var}{\mathrm{Var}}

\newcommand{\Cov}{\mathrm{Cov}}
\newcommand{\normltwo}{L^2}
\newcommand{\normlp}{L^p}

\newcommand{\parents}{Pa} % See usage in notation.tex. Chosen to match Daphne's book.

\usepackage{arydshln}
\usepackage{xcolor}

\usepackage{hyperref}
\usepackage{url}

\usepackage{graphicx}
\usepackage{caption}
\usepackage{subcaption}

\usepackage{amssymb}
\usepackage{xcolor}

\title{Low-rank projections of GCNs Laplacian}
\iclrfinalcopy
\author{Nathan Grinsztajn  \\
Inria, Univ. Lille, CNRS\\
Lille, France\\
\texttt{nathan.grinsztajn@inria.fr} \\
\And
Philippe Preux \\
Inria, Univ. Lille, CNRS\\
Lille, France\\
\texttt{philippe.preux@inria.fr} \\
\And
Edouard Oyallon \\
CNRS, LIP6, Sorbonne University\\
Paris, France\\
\texttt{edouard.oyallon@lip6.fr}
}

\usepackage{todonotes}
\begin{document}

\maketitle

%Our finding is that a GCN by design mainly uses the low frequencies, and that the role of the high frequencies contrast with their standard use in signal processing. Here, we are able to design MLPs that rely simply on a few eigenvectors of the graph Laplacian while being competitive with supervised graph approaches.

\begin{abstract}
In this work, we study the behavior of standard models for community detection under spectral manipulations. Through various ablation experiments, we evaluate the impact of bandpass filtering on the performance of a GCN: we empirically show that most of the necessary and used information for nodes classification is contained in the low-frequency domain, and thus contrary to images,  high frequencies are less crucial to community detection. In particular, it is sometimes possible to obtain accuracies at a state-of-the-art level with simple classifiers that rely only on a few low frequencies.
% this is surprising because contrary to GCNs, no cascade of filtering along the graph structure is involved and it indicates that the important spectral components for the supervised community detection task are essentially in the low-frequency domain. 

% Nathan{ We further derive a simple method to adjust the trade-off between smoothing and high-frequency propagation for deeper GCNs. It involves no explicit Laplacian diagonalization and consistently boosts performances on standard and challenging benchmarks.}

\end{abstract}

\section{Introduction}

Graph Convolutional Networks (GCNs) are the state of the art in community detection~\citep{kipf2016semi}. They correspond to Graph Neural Networks (GNNs) that propagate graph features through a cascade of linear operator and non-linearities, while exploiting the graph structure through a linear smoothing operator. However, the principles that allow GCNs to obtain good performances remain unclear.
% It is suggested in \cite{li2018deeper} that GCNs are eager to over-smooth their representation, which indicates they average too much neighborhood nodes and dilute classification information. The smoothing is generally interpreted as a low-pass filtering through the graph Laplacian, and finding a way to exploit high frequencies of the graph Laplacian is an active research question~\citep{oono2019graph}.
% In contrast to this, 
Our work actually suggests that, in the setting of community detection, graph Laplacian high frequencies have actually a minor impact on the classification performances of a standard GCN, as opposed to standard Convolutional Neural Networks for vision.
% , which are built thanks to image processing considerations.

% Graph Signal Processing (GSP) is a popular field whose objective is to manipulate signals spectrum whose topology is given by a graph. Typically, this graph has a non-Euclidean structure, however many central theoretical results \citep{hammond2011wavelets} are based on an analogy with Euclidean, regular grids. For instance, a spectral component or frequency has to be understood as an eigenvalue of the Laplacian, yet it thus suffers from intrinsic issues such as isotropy~\citep{icml2020_2411}. The principles of GSP are very appealing because they allow to use the dense literature of harmonic analysis, on graphs. Thus, this literature is at the core of many intuitions and drives many key ingredients of a GCN design, which evokes standard tools of signal processing: convolutions, shift invariance, wavelets, Fourier~\citep{bronstein2017geometric}, etc. Here, we certainly observe several limits of this analogy in the context of community detection: for instance, we observe that discarding high frequencies has a minor impact on a GCN behavior, because the spectrum of the graphs of the datasets that are used is essentially located in the low-frequency domain. This type of ideas is for instance core in spectral clustering algorithms.

% \Edouard{Je mergerai avec le paragraphe précédent, quitte à ne PAS parler de SBM et Fielder vector mais juste du fait que les gens aiment le laplacien}
Spectral clustering is a rather different point of view from deep supervised GCNs which studies node labeling in unsupervised contexts: it generally relies on generative models based on the graph spectrum.
% The main principle is to consider the eigenvectors corresponding to the smallest non-zero eigenvalues, referred to as Fiedler vectors~\citep{doshi2020fiedler}: those directions allow to define clusters, depending on the sign of a feature.
% Several theoretical guarantees can be obtained in the context of Stochastic Block Model approximation ~\citep{rohe2011spectral}. 
Our paper proposes to establish a  clear link between these approaches:  we show that the informative graph features are located in a low-frequency band of the graph Laplacian and can be efficiently used in a deep supervised classifier.% such as .

This paper shows that experiments on standard community detection datasets like Cora, Citeseer, Pubmed can be conducted using only few frequencies of their respective graph spectrum without observing any significant performances drop. Other contributions are as follows: \textbf{(a)} First we show that most of the necessary information exploited by a  GCN for a community detection task can actually be isolated in the very first eigenvectors of a Laplacian.
% \textbf{(b)} We numerically show that the high-frequency eigenvalues are less informative for the supervised community detection task and that a trained GCN is more stable to them.
\textbf{(b)} We observe that a simple Multi-Layer Perceptron (MLP) method fed with handcrafted features accounting for low-frequency eigenvalues allows to successfully deal with transdusctive datasets like Cora, Citeseer or Pubmed.
% : to our knowledge, this is the first competitive results obtained with a MLP on those datasets.

% We now discuss the organization of the paper: first, we discuss the related work in Sec.~\ref{related}. We explain our notations as well as our work hypotheses in Sec.~\ref{hypo}. Then, we study low-rank approximations of the graph Laplacian in Sec.~\ref{low}. 
% Finally, the end of Sec.~\ref{hyperparams} proposes several experiments to study the impact of high frequencies on GCNs. A basic code is provided in the supplementary materials, and our code will be released on an online public repository at the time of publication.

\section{Related Work}
% \Edouard{Combine avec l'intro - je suggère de fusionner des passages et de retirer les "another line of work", mais de quand même citer les travaux - retire 2 related work pas utile et compress le reste, c pas (si) important le related work dans un workshop paper}
\label{related}

\paragraph{GCNs and Spectral GCNs} Introduced in \citet{kipf2016semi}, GCNs allow to deal with large graph structure in semi-supervised classification contexts. This type of model works at the node level, meaning that it uses locally the adjacency matrix. This approach has inspired a wide range of models, such as linear GCN \citep{wu2019simplifying}, Graph Attention Networks \citep{velivckovic2017graph}, GraphSAGE \citep{hamilton2017inductive}, etc. In general, this line of work does not consider directly the graph Laplacian. Another line of work corresponds to spectral methods, taking inspiration from Graph Signal Processing (GSP), a popular field whose objective is to manipulate signals spectrum whose topology is given by a graph. They employ filters which are designed from the spectrum of a graph Laplacian~\citep{bruna2013spectral,icml2020_2411,oyallon2017analyzing,grinsztajn2021inter}. In general, those works make use of polynomials in the Laplacian ~\citep{defferrard2016convolutional}
% , which are very similar to an anisotropic diffusion~\citep{klicpera2019diffusion}
. All those references share the idea to manipulate bandpass filters that discriminate the different ranges of frequencies.

\paragraph{Over-smoothing in GCNs} In the context of GCN, \citet{li2018deeper} is one of the first papers to notice that cascading low-pass filters can lead to a substantial information loss. The result of our work indicates that the important spectral components for detecting communities are already in the low-frequency domain and that this is not due to an architecture bias. \citet{zhao2019pairnorm,yang2020revisiting} propose to introduce regularizations which address the loss of information issues. \citet{cai2020note, oono2019graph} study the spectrum of a graph Laplacian under various transforms, yet they consider the spectrum globally and in asymptotic settings, with a deep cascade of layers. \citet{huang2020tackling,rong2019dropedge} introduce data augmentations, whose aim is to alleviate over-smoothing in deep networks: we study GCNs without this ad-hoc procedure.

\paragraph{Spectral clustering and low rank approximation} As the literature about spectral clustering is large, we mainly focus on the subset that connects directly with GCN. \citet{mehta2019stochastic} proposes to learn an unsupervised auto-encoder in the framework of a Stochastic Block Model. \citet{oono2019graph} introduces the Erdös – Renyi model in the GCN analysis, but only in an asymptotic setting. \citet{loukas2018spectrally} studies the graph topology preservation under the coarsening of the graph, which could be a potential direction for future works. \citet{liao2018lanczosnet} uses the Lanczos algorithm to construct low rank approximations of the graph Laplacian, which facilitates efficient computation of matrix powers.

\paragraph{Node embedding} A MLP approach can be understood as an embedding at the node level. For instance, \citet{aubry2011wave} applies a spectral embedding combined with a diffusion process for shape analysis, which allows point-wise comparisons. We also mention \citet{deutsch2020spectral} that uses a node embedding, based on the spectrum of a quite modified graph Laplacian, obtained from on a measure of node centrality.

% \paragraph{Graph Scattering Networks (GSN)} This class model explicitly employs band-pass based on the spectrum of a graph Laplacian and it is thus necessary to review it. \citet{gao2019geometric,gama2018diffusion,gama2019stability} are a class of neural networks built upon an analogy with a Scattering Transform \citep{mallat2012group}. They typically rely on a cascade of wavelets followed by an absolute value: the objective of each wavelet is to separate multi-scale information into dyadic bandpass filters. This method relies heavily on each eigenvector of the Laplacian and the dyadic space is typically constructed from a diffusion process at dyadic intervals. Interferometric Graph Transform on the other hand relies on the concept of demodulation, which is clear in the context of Lie groups but unclear for community detection tasks \citep{icml2020_2411}.

\paragraph{GCN stability} Stability of GCNs has been theoretically studied in \citet{gama2019stability}, which shows that defining a generic notion of deformations is difficult. Surprisingly, it was noted in \citet{icml2020_2411} that stability is not a key component to good performances. The stability of GCN has also been investigated in \citet{verma2019stability} but only considers neural networks with a single layer and relies on the whole spectrum of the learned layer. \citet{keriven2020convergence} considers the stability of GCNs, and relies on an implicit Euclidean structure: it is unclear if this holds in our settings. \citet{sun2020fisher} is one of the first works to study adversarial examples linked to the node connectivity and introduces a loss to reduce their effects. \citet{zhu2019robust} also addresses the stability issues by embedding GCNs in a continuous representation. None of these work directly related a trained GCN to spectral perturbations.

\section{Framework}\label{hypo}
\subsection{Method}
We first describe our baseline model. Our initial graph data are a graph with $N$ nodes, its adjacency matrix $A$ with diagonal degree matrix $D$, and node features $X$. We consider first-order GCNs as introduced in \citep{kipf2016semi}, which correspond to GNNs that propagate node feature input $H^{(0)}\triangleq X$ through a cascade of layers, via the iteration:
\begin{equation}
H^{(l+1)}\triangleq\sigma\bigg( \tilde AH^{(l)}W^{(l)}\bigg)\,, \label{recur}
\end{equation}

where $\tilde A=\frac{1}{2}(I+D^{-1/2}AD^{-1/2})$, $\sigma$ a point-wise non-linearity and $W^{(l)}$ a parametrized affine operator. Note that if $\tilde A=I_N$, then Eq. \ref{recur} is simply an MLP. The $\frac 12$ factor is a normalization factor to obtain $\Vert\tilde A\Vert=1$. In the semi-supervised setting, a final layer $f(X,\tilde A)\triangleq H^{(L)}$ is fed to a supervised loss $\ell$ (here a softmax) and $\{W^{(0)},...,W^{L-1}\}$ are trained in an end-to-end manner to adjust the label of each node. We note that for undirected graph, $\tilde A$ is a positive definite matrix with positive weights, which can be understood as an averaging operator \citep{li2018deeper}.
% , as, ignoring $\tilde D$, we see that for some node features $X$, we have:
% \begin{equation}
% [(I_N+A)X]_i= X_i+\sum_{j\rightarrow i}A_{i,j}X_j\,.
% \end{equation}

% We remark that multiple choices of averaging operators are possible: as briefly discussed in Appendix \ref{GCNvsAugNorm_appendix} other formulations did not change our numerical conclusions, thus we decided to keep the simplest to be handled mathematically. 
We are interested in analyzing the properties of spectral approximations of $\tilde A$. We consider the decreasing set of eigenvalues $\Lambda=\{\lambda_k\}_{k\geq 0}$ of $\tilde A$, such that $\lambda_{k}\geq \lambda_{k+1}$, and we denote by $u_k$ the $k$-th eigenvector corresponding to $\lambda_k\in\Lambda$. We remind that $\Lambda \subset [0,1]$ and that the eigenspace corresponding to $\lambda=1$ can be interpreted as the kernel of the symmetric normalized graph Laplacian. We then write $\tilde A_{[k_1,k_2]}\triangleq\sum_{k_1\leq k\leq k_2} \lambda_k u_ku_k^T$
% \begin{equation}
% \tilde A_{[k_1,k_2]}\triangleq\sum_{k_1\leq k\leq k_2} \lambda_k u_ku_k^T\,,
% \end{equation}
, such that $\tilde A=\tilde A_{[0,N]}$. We are interested in studying the degradation accuracy if we replace $\tilde A$ with $\tilde A_{[0,k]}$ or $\tilde A_{[k,N]}$ for some $0<k<N$.
% The next section explains that under standard but oversimplifying assumptions, an approximation of the type $\tilde A_{[0,k]}$ is relevant for community detection tasks.

\section{Experiments}
\label{hyperparams}

Matching the previous work practice, we focus on the three classical benchmark dataset for community detection: Cora, Citeseer and Pubmed \citep{cora_et_al}. The task consists in classifying the research topic of papers in three citation datasets. Those tasks are transductive, meaning all node features are accessible during training. We use the same training/validation/test split as \citet{dropEdge_full1}, \citet{dropedge_full2}, and \citet{rong2019dropedge} on all datasets in our experiments, which corresponds to a supervised learning scenario.
% Fig. \ref{spectrum} plots the 3 spectra $\Lambda$ in decreasing order for each of those datasets. The three datasets exhibit a significant spectral gap, which is aligned with the model of Sec. \ref{}.
% Note that Pubmed has one connected component, and that the decay of its spectrum is fast compared to Cora or Citeseer, which indicates a low-dimensional structure~\citep{belkin2002laplacian}.
The statistics of each dataset are listed in the supplemental materials, as well as the training details.

%We follow the standard full-supervised training procedure of \citet{dropedge}. 

%Similarly to \cite{kipf2016semi}, we consider the datasets Cora, Citeseer, Pubmed and ???. We followed the training procedure of \citep{}. Each dataset is ...

\subsection{Low rank approximation}\label{low}

%\subsubsection{Fiedler vectors and GCNs}

\begin{figure}[h!]
  \centerline{\includegraphics[width=0.7 \linewidth]{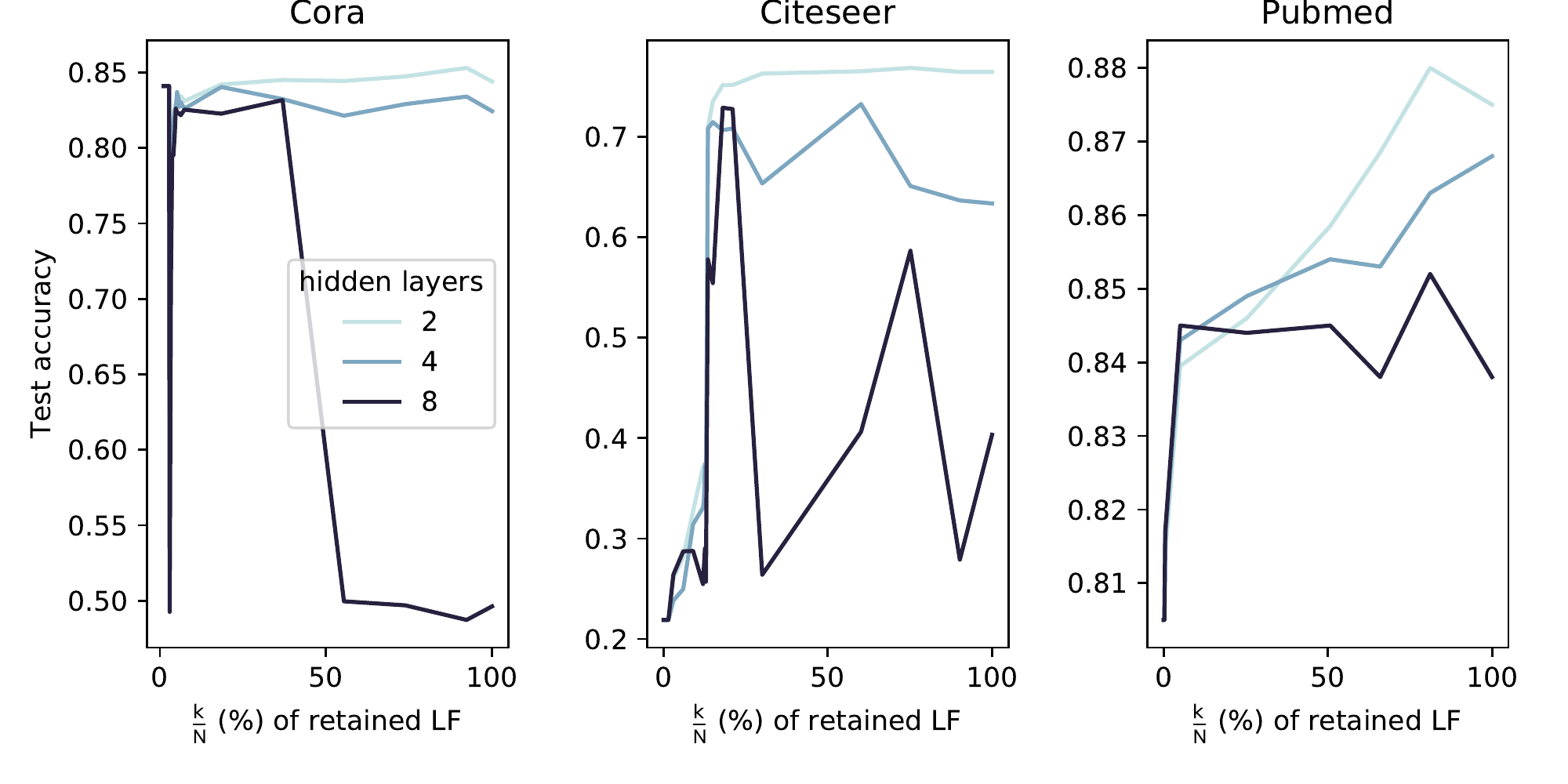}}
  \caption{Test accuracies reached by a GCN as a function of the Low-Frequency (LF) band $[\lambda_{k},\lambda_0]$ retained by $\tilde A_{[0,k]}$, for various depths (100$\%$ corresponds to the full spectrum, including low frequencies). This figure indicates that the informative components are located in the LF domain.}
  \label{testAccNoL}
\end{figure}

\begin{figure}[h!]
  \centerline{\includegraphics[width=0.7 \linewidth]{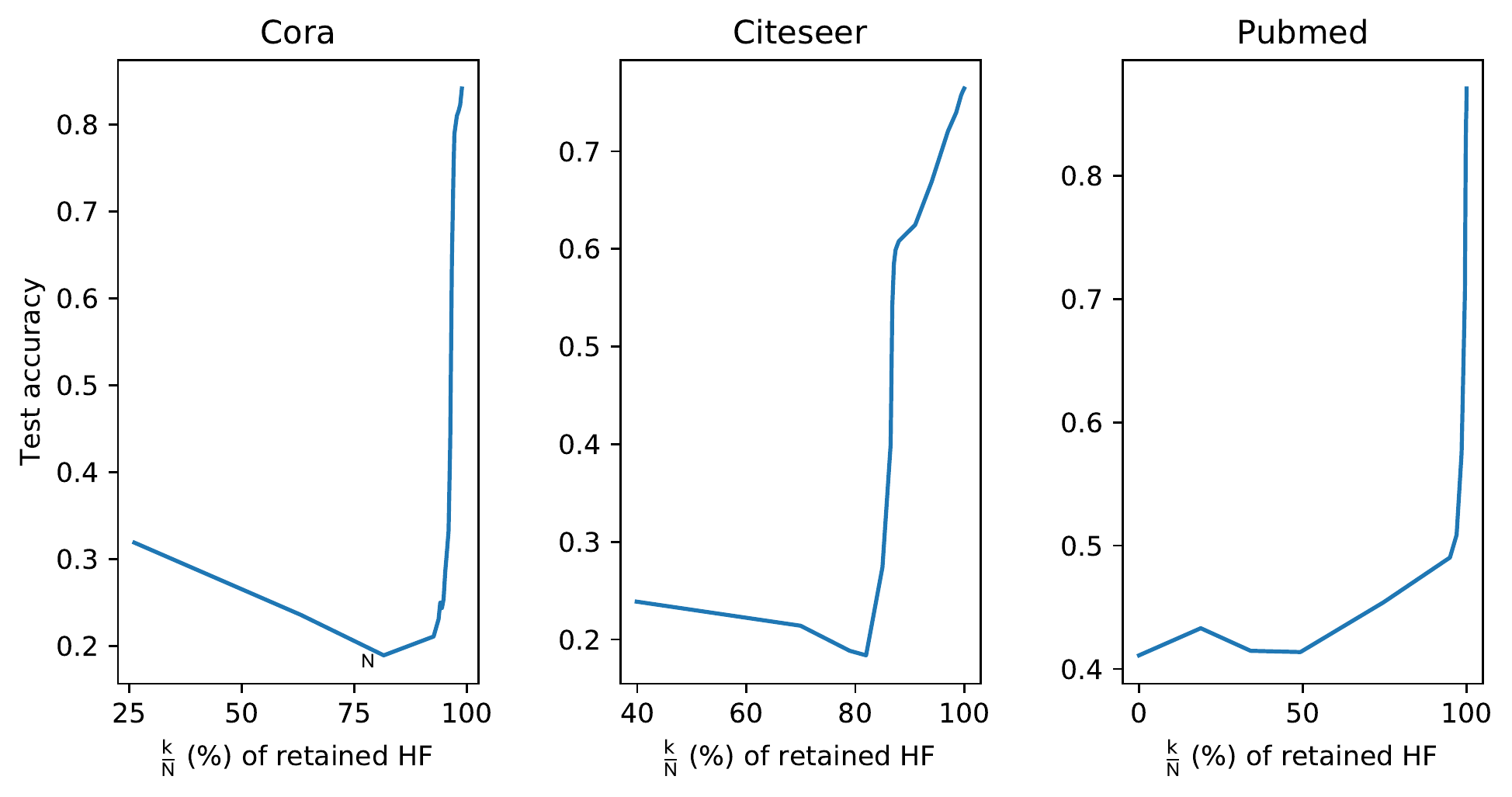}}
  \caption{Test accuracies reached by a GCN as a function of the high frequencies (HF) band $[\lambda_{N},\lambda_{N-k}]$ retained by $\tilde A_{[N-k,N]}$, for a GCN of depth 2  (100$\%$ corresponds to the full spectrum). This figure indicates that high frequencies are less informative for a community detection task.}
  \label{testAccNoH}
\end{figure}

\paragraph{GCN ablation}Here, we consider the two projections $\tilde A_{[0,k]}$ and $\tilde A_{[N-k,N]}$, where $k$ is adjusted to retain only a portion of the spectrum. Those projections can be interpreted respectively as high-pass and low-pass filters. As detailed in Appendix \ref{understandlowrank}, those projections will allow to study which frequency band is important for the community detection task. Fig. \ref{testAccNoL} and Fig. \ref{testAccNoH} report the respective performance when considering the models $f(X,\tilde A_{[N-k,N]})$ and $f(X,\tilde A_{[0,k]})$ for some $k$.
On Fig.~\ref{testAccNoL}, we observe that retaining only very few frequencies (less than 10\%) does not degrade much the accuracy of the original network. This does not contradict the observation of \cite{li2018deeper} which studies empirically the over-smoothing phenomenon, as our finding indicates that a GCN uses mainly the low frequency domain. For Pubmed, using almost all the high frequencies is required to recover the original accuracy of our model. Interestingly, deeper GCNs seem to benefit from the high frequency ablation, yet their accuracy remains below their shallow counter-part and they are still difficult to train. This instability to spectral perturbations is further studied in Appendix \ref{grad}. Fig. \ref{testAccNoH} indicates that the major information for supervised community detection is contained in the low frequencies: dropping the latter leads to substantial accuracy drop, even for a shallow GCN.

\begin{figure}[h!]
  \centerline{\includegraphics[width=0.7 \linewidth]{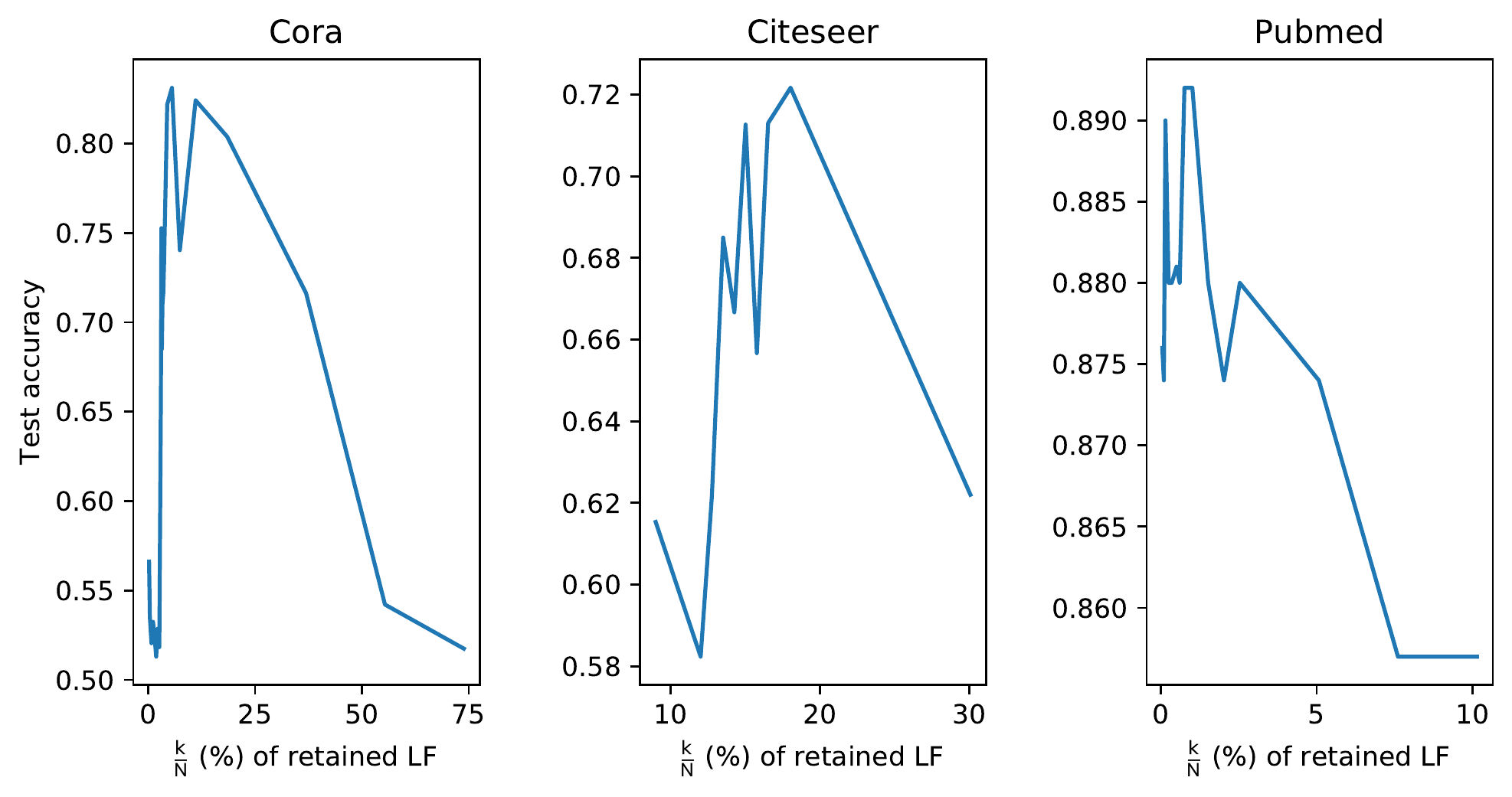}}
\caption{Accuracy of MLPs trained on $ X_k$, according to the low frequency band $[\lambda_k,\lambda_0]$ retained. We note that selecting a narrow low-frequency band can lead to competitive accuracies.}
\label{mlp}
\end{figure}

\paragraph{MLP augmentation} We further study the importance of low frequencies via a spare ablation experiment based on a MLP. We augment each graph features $X$ through the concatenation $ X_k=[X,u_1^T,...,u_k^T]$ of the first $k$ eigenvectors.
% For a fair evaluation, we used exactly the same hyper-parameters as for the experiments above.
Fig. \ref{mlp} reports the accuracy of our MLP trained on $X_k$ as a function of $k$. We see that using a fraction $\frac{k}{N}\leq 20\%$ of the eigenvectors allows to recover the performance of a GCN. More surprisingly, we note that as $k$ increases, the accuracy drops, which indicates that high frequencies behave like a residual noise that is overfitted. This experiment emphasizes that a low rank approximation of the Laplacian can be beneficial to a MLP classifier.

\begin{table}[th!]
\small
\caption{Comparison of various models on Cora, Citeseer and Pubmed. \label{acc}}
\label{sample-table}
\begin{center}
\begin{tabular}{lcccc}
\multicolumn{1}{c}{\bf Method}  &\multicolumn{1}{c}{\bf Data augmentation}&\multicolumn{1}{c}{\bf Cora}&\multicolumn{1}{c}{\bf Citeseer}&\multicolumn{1}{c}{\bf Pubmed}
\\ \hline \\
GCN \citep{rong2019dropedge} &No& \textbf{86.6} & \textbf{79.3} & 90.2 \\
Fastgcn \citep{dropedge_full2}               &No&86.5&-&88.8 \\
MLP on $X$     & No & 74.0 & 73.3 & 89.1\\
MLP on $\tilde X_k$ (ours) &No&\textbf{86.6}&77.3&\textbf{91.4}\\
\hdashline\\
DropEdge   \citep{rong2019dropedge}       &Yes&88.2&80.5&91.7 \\
\citep{huang2018adaptive}          &Yes&87.4&79.7&90.6 \\
\end{tabular}
\end{center}
\vspace{-10pt}
\end{table}

\paragraph{Boosting MLP performances} We perform a hyper-parameter grid search on each dataset to investigate MLP strengths further. We summarize our findings in Tab. \ref{acc}. In particular, one uses a fraction 6\%, 15.0\% and 0.7\% of the spectrum respectively on Cora, Citeseer and Pubmed in order to obtain our best performances. 
% We note that a MLP trained solely on $X$ already outperforms the approach of \cite{dropedge_full2}, without relying on a graph structure. 
More details on the methodology are provided in the supplementary material. This simple model is often competitive with concurrent works, and in particular with vanilla GCNs. Note also that our method does not incorporate any data augmentation procedure. 
% We could also potentially incorporated data augmentation procedures, at the price of an extended computation time. 
We conclude that GCNs do not compute more complex invariants than a MLP fed with low frequencies, in the context of community detection.

\section{Conclusion}
% \Edouard{1 seul paragraphe}
We have studied the classification performance of a GCN under low-pass and high-pass filtering, in the context of community detection. Our finding is that GCNs   rely significantly on the low frequencies, and can even benefit from high-frequency ablations. Then, we are able to design MLPs that rely simply on a few eigenvectors of the graph Laplacian and can be competitive with GCNs. 
% We also study the stability of a GCN w.r.t. spectral perturbations, and show that they are more robust to high-frequency, which is counter-intuitive when compared to vanilla CNNs.

% Our work indicates that not only more difficult graph benchmarks are necessary, but also benchmarks whose optimal model would rely on the high frequencies of a graph Laplacian. 
%It also shows that standard GSP tools such as graph wavelets might be simplified to low-pass filters, in the context of community detection, as 
We observed that the high-frequency does not bring a significant amount of information, and can even be interpreted as a residual noise in this particular setting. Our methodology can also help to identify if the accuracy improvement of a new  architecture is due to a better processing of high frequencies.
\subsection*{Acknowledgements}
EO was granted access to the HPC resources of IDRIS under the allocation 2020-[AD011011216R1] made by GENCI and this work was partially supported by ANR-19-CHIA "SCAI". Experiments presented in this paper were partially carried out using the Grid'5000 testbed, supported by a scientific interest group hosted by Inria and including CNRS, RENATER and several Universities as well as other organizations. NG is recipient of a PhD funding from AMX program, Ecole polytechnique.
\newpage

\bibliography{iclr2021_conference}
\bibliographystyle{iclr2021_conference}
\newpage
\appendix
\section{Appendix}
\subsection{Dataset statistics}

\begin{table}[th]
\caption{Dataset Statistics}
\label{dataset_statistics}
\begin{center}
\begin{tabular}{cccccc}
\bf Datasets & \bf Nodes & \bf Edges & \bf Classes & \bf Features & \bf Traing/Validation/Testing split
\\ \hline \\
Cora & 2,708 & 5,429 & 7 & 1,433 & 1,208/500/1,000 \\
Citeseer & 3,327 & 4,732 & 6 & 3,703 & 1,812/500/1,000 \\
Pubmed & 19,717 & 44,338 & 3 & 500 & 18,217/500/1,000 \\
\end{tabular}
\end{center}
\end{table}

\subsection{Spectra of Cora, Citeseer, Pubmed}

\begin{figure}[h!]
  \centerline{\includegraphics[width=0.7 \linewidth]{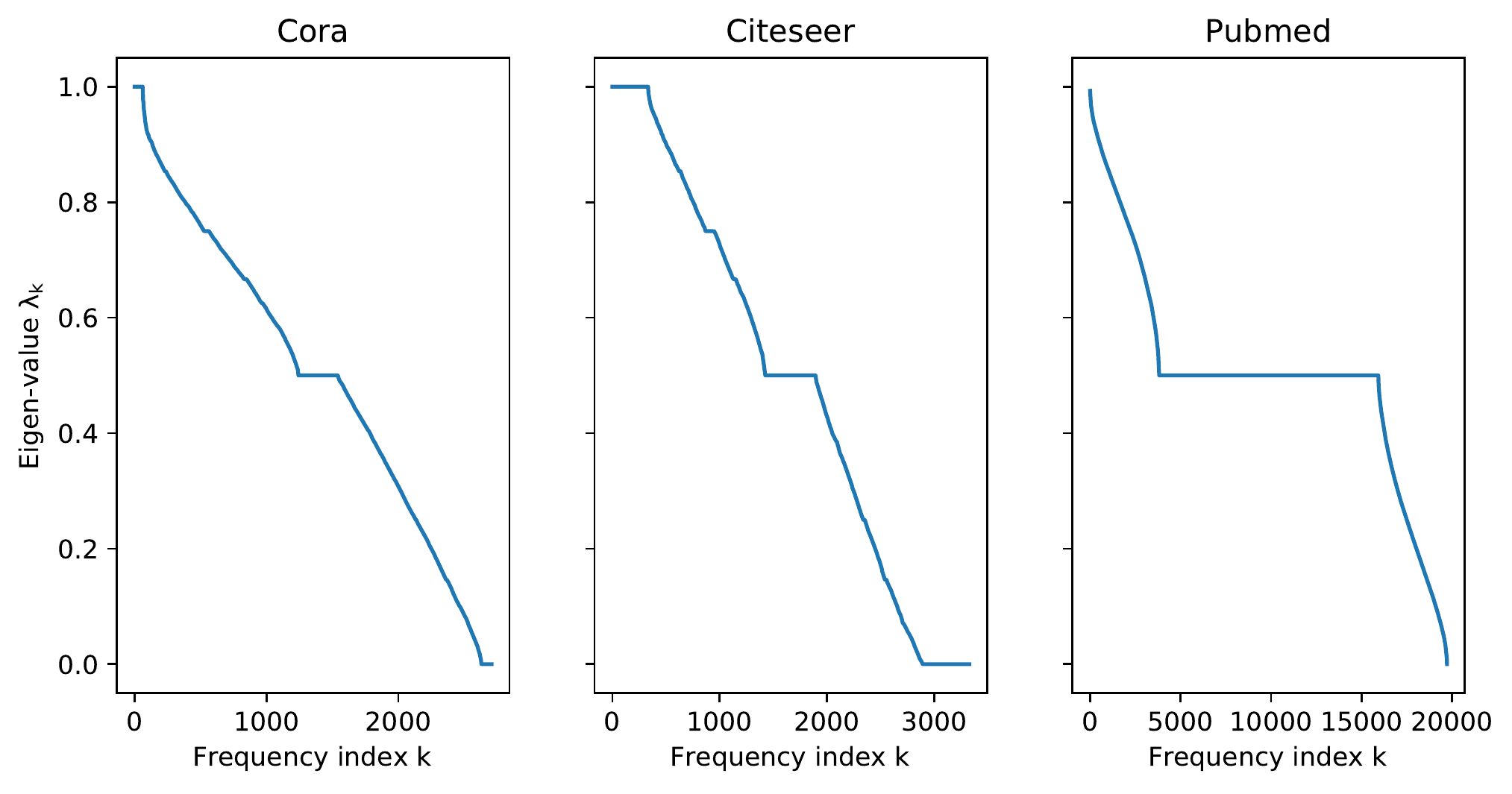}}
  \label{spectrum}
  \caption{Spectrum $\Lambda$ of $\tilde A$. Note that the eigenvalues are decreasing, displayed with multiplicity. Observe that the decay of the spectrum is fast. \label{spectrum}}
\end{figure}

\section{Training Details}

We choose $\sigma$ to be the ReLU non-linearity \citep{krizhevsky2012imagenet}. Unless specified otherwise, the weights of our models (either GCN or MLP) are optimized via Adam, with an initial learning rate 0.01 and weight decay of 0.001, during 800 epochs. We use by default a dropout of 0.5. Our model consists in GCN layers with 2 hidden layers of size 128. When using grid search, we cross validate our hyper-parameters on a validation set, using early stopping with a patience of 400. The grid search is detailed in the following table. Unless specified otherwise, each plot is obtained by an average over 3 different seeds.

\subsection{Hyper-parameter Description}

\begin{table}[th]
\small
\caption{Hyper-parameter Description and Grid Search Range}
\label{table_hyp}
\begin{center}
\begin{tabular}{ccc}
\multicolumn{1}{c}{\bf Hyper-parameter}  &\multicolumn{1}{c}{\bf  Description}
&\multicolumn{1}{c}{\bf  Range}
\\ \hline \\
lr & learning rate & $\{ 0.01, 0.001 \}$ \\
hidden layers & the number of hidden layers & $\{1, 2, 3, 4\}$ \\
weight-decay & L2 regulation weight & 0.001 \\
dropout & dropout rate & $\{ 0, 0.2, 0.4, 0.6, 0.8\}$\\
frequencies & the number of low frequencies to add (\%) & $\{ 0.1, 0.2, 0.4, 0.7, 1, 2, 3,$ \\ &&$6, 10, 15, 20, 30, 50, 80, 100\}$
\\
eigenvector features normalization & whether to normalize the new MLP features & \{ False, True \} \\ & in line (per nodes) or column (per features) & 

\end{tabular}
\end{center}
\end{table}

% \section{On the choice of the normalization}
% \label{GCNvsAugNorm_appendix}
% GCNs defined in \citet{kipf2016semi} don't exactly used the first order Laplacian approximation, but introduce a normalization trick : $\tilde A = (D+I_N)^{-\frac{1}{2}} (A+I_N) (D+I_N)^{-\frac{1}{2}} $. Using the standard hyper-parameters defined in Section \ref{hyperparams}, we averaged our results on 3 seeds and observed no significant difference across datasets, as shown in Fig. \ref{GCNvsAugNorm}.

% \begin{figure}[h]
%   \centerline{\includegraphics[width=1.1 \linewidth]{figures/rez_vs_regularGCN.pdf}}
%   \caption{Comparison between the standard Laplacian first order normalization and the GCN normalization trick}
%   \label{GCNvsAugNorm}
% \end{figure}

\section{Understanding low rank approximations for GCNs}
\label{understandlowrank}
\begin{figure}[h!]
     \centering
     \begin{subfigure}[b]{0.3\linewidth}
         \centering
         \includegraphics[width=\linewidth]{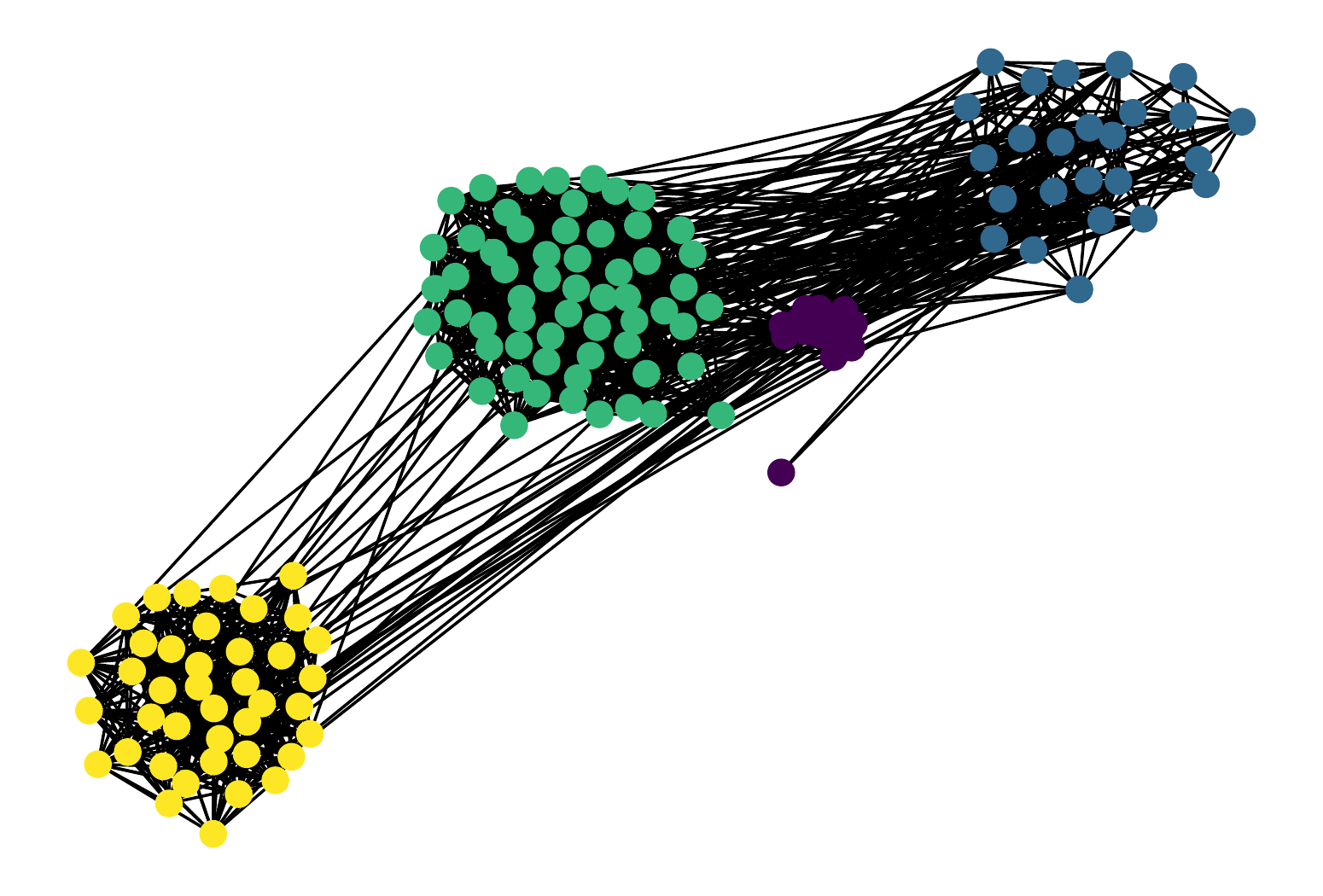}
         \caption{}
     \end{subfigure}
     \hfill
     \begin{subfigure}[b]{0.6\linewidth}
         \centering
         \includegraphics[width=\linewidth]{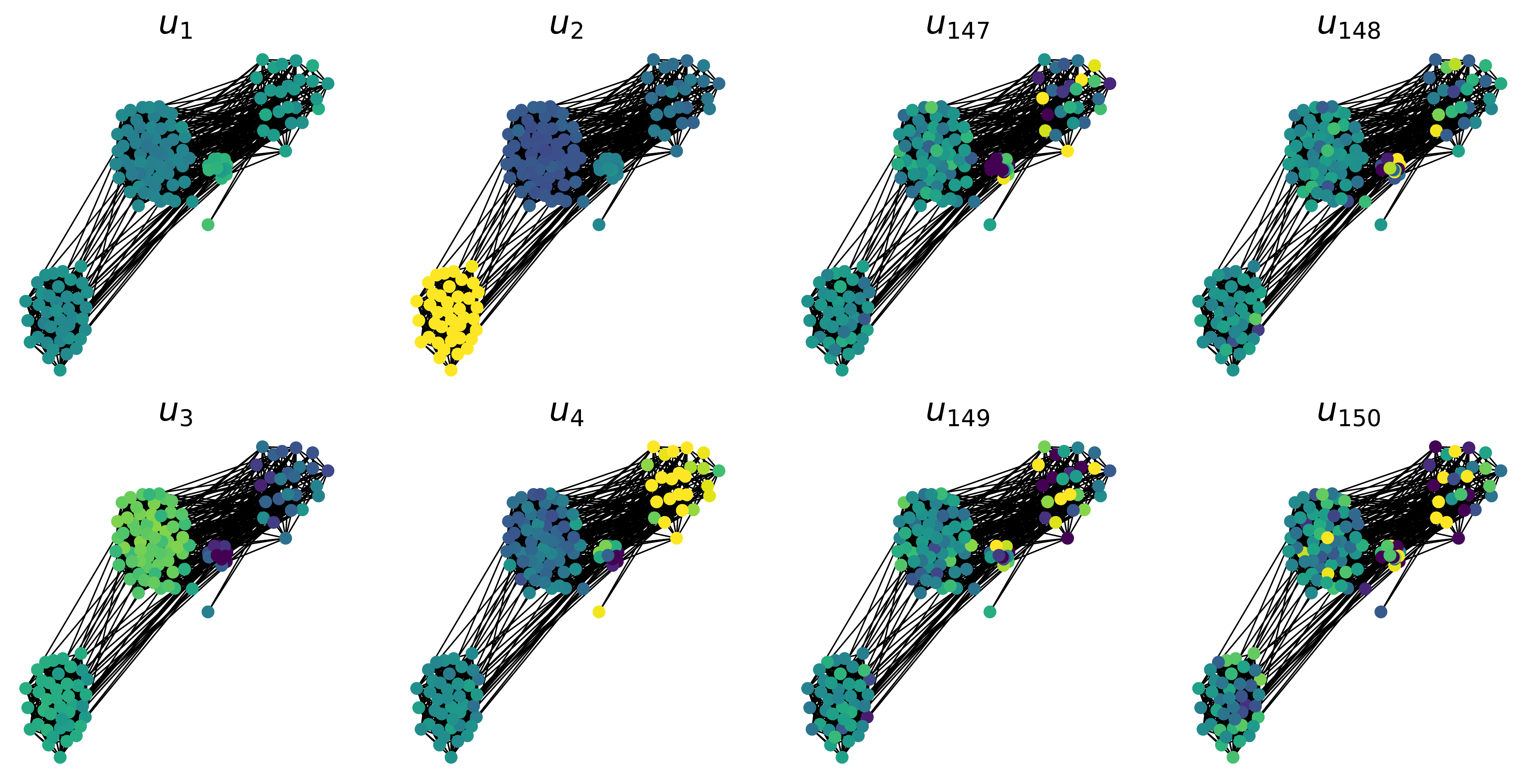}
         \caption{}
         \label{}
     \end{subfigure}
        \caption{Under a Stochastic Block Model with 4 communities (a), we represent the first Eigenvectors (b, left) and the last Eigenvectors (b, right) of $\tilde A$. On the left figure, the colors stand for the communities. On the right, they stand for the values of the considered eigenvectors (the brighter the higher). A low rank approximation maintains the information related to the different communities.}
        \label{fig:SBM}
\end{figure}

We now justify our approach under the standard setting of the Stochastic Block Model \citep{abbe2017community}, and we will simply remind the reader of several elementary results. This model corresponds to a generative model that describes the interaction between $r$ communities $\{C_1,...,C_r\}$. Assuming that two nodes $i,j$ belong to the communities $C_{r_i},C_{r_j}$, an edge is sampled with a probability $p_{r_i,r_j}$ \citep{rohe2011spectral}. For the sake of simplicity, let us assume that $r=2$, that the probability of an edge between two nodes $i,j$ is $p$ if those nodes belong to the same community and $q<p$ otherwise, and that both communities correspond to $|C|$ nodes. In this case, the unnormalized expected adjacency matrix is given by:
\begin{equation}
\begin{bmatrix} p&\cdots &p&q&\cdots&q \\
\vdots& &\vdots&\vdots&&\vdots\\
p&\cdots &p&q&\cdots&q\\
q&\cdots &q&p&\cdots&p \\
\vdots& &\vdots&\vdots&&\vdots\\
q&\cdots &q&p&\cdots&p\\
\end{bmatrix}\,,
\end{equation}

where we grouped in matrix block the nodes from the same community. We note that the two dominant eigenvectors are given by:
\begin{equation}
    u_1=[\overbrace{1, ..., 1,}^{|C|\text{ times}}\underbrace{1, ..., 1}_{|C|\text{ times}}]\text{ and }u_2=[\overbrace{1, ..., 1, }^{|C|\text{ times}}\underbrace{-1, ..., -1}_{|C|\text{ times}}]\,.
\end{equation}
 Observe that the second eigenvector $u_2$ captures all the information about the two communities, through the sign of its coefficients. Here, the spectral gap (the ratio between the two dominant eigenvalues) is given by $0\leq \frac{p-q}{p+q}<1$ and ideally this spectral gap should be as large as possible for identifying the two communities. If the number of nodes is large, concentration results \citep{wainwright2019high} imply that the empirical adjacency matrix concentrates around its expectation, and that under this assumption, a low-rank approximation $\tilde A_{[0,2]}$ captures most of the available information about the two communities. We illustrate this idea on Fig. \ref{fig:SBM}. While these assumptions might not hold in practice, it justifies why low-rank approximations of a Laplacian are relevant in the setting of community detection and it explains why high frequencies might not be as important as low frequencies for supervised community detection task. The next section validates empirically this approach in the context of GCNs and simpler architectures.

\section{Stability to High Frequencies}
\label{grad}
\begin{figure}[h]
  \centerline{\includegraphics[width=1.1 \linewidth]{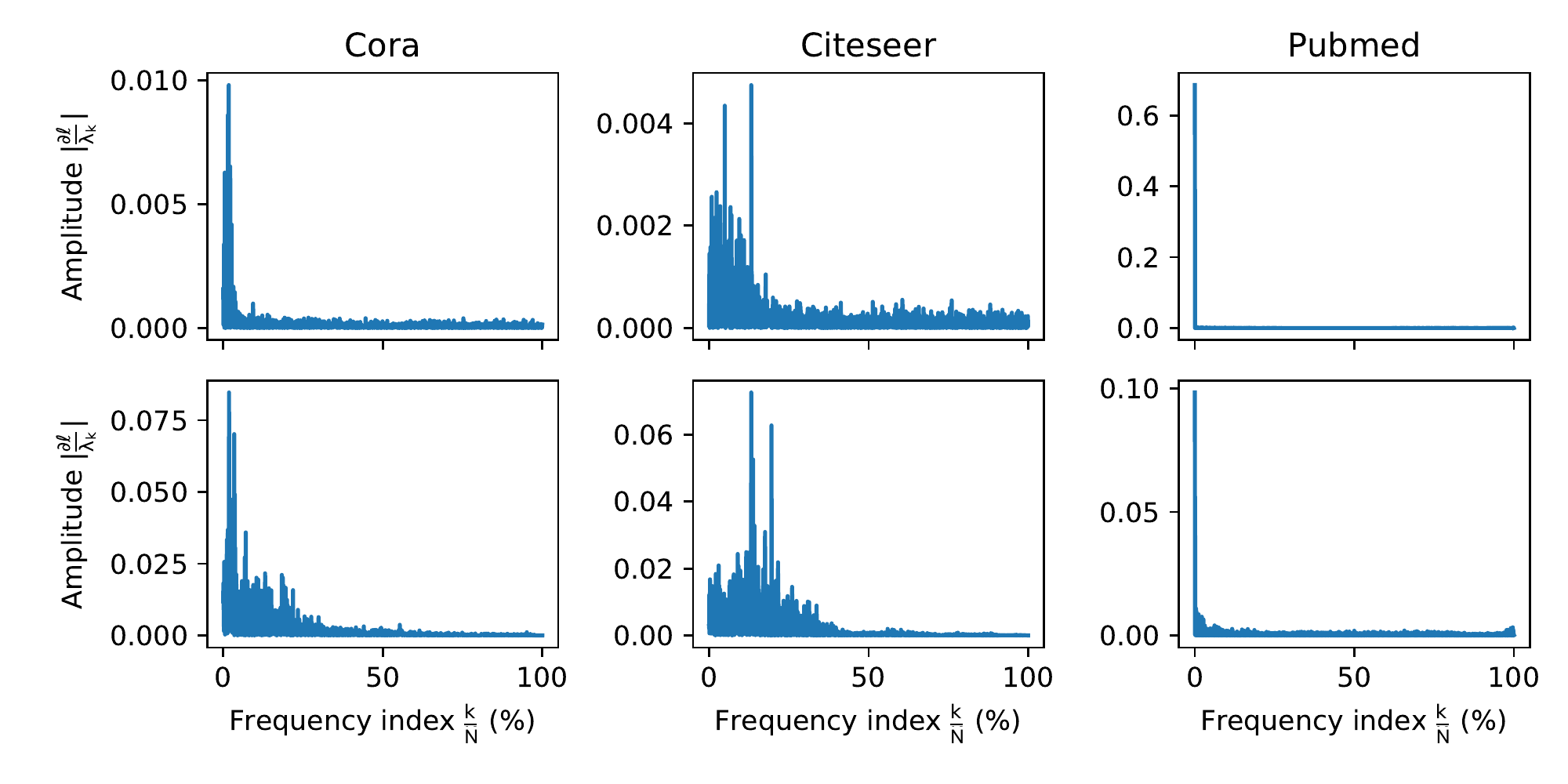}}
  \caption{The (top) and (bottom) figures corresponds to $|\frac{\partial \ell}{\lambda_k}|$ of the same model taken respectively at the initialization and the end of training. Note that high frequencies are more stable than low frequencies for the three datasets.}
  \label{spectral_grad_full}
\end{figure}

We now study the stability of a GCN w.r.t.\@ spectrum perturbations. In the case of image processing, it is standard \cite{} that low frequencies almost do not affect the classification performances and that perturbations of high frequencies lead to instabilities. We would like to validate that this principle does not hold here, and to do so, we consider $\nabla_{\Lambda} \ell$, which is the gradient w.r.t.\@ every singular value $\lambda_k$. Small amplitudes of $|\frac{\partial \ell}{\partial \lambda_k}|$ indicate more stable coefficients. Fig.~\ref{spectral_grad_full} plots the amplitude of the gradient w.r.t.\@ $\lambda_k$ at the initialization of a GCN and after training. First, we note that a GCN is more sensitive to spectral perturbations after training, which is logical because the GCN adapts its weights to the specific structure of a given graph. After training, we remark that the high-frequency perturbations have a small impact compared to the low-frequency perturbations on the three datasets, except for Pubmed which has dominant low frequencies. This is consistent with our previous findings.

\section{Note on the computational overhead}
The MLPs introduced above are of interest if the corresponding graph topology is fixed, with a large graph, and high connectivity. Indeed, using a MLP allows to easily employ mini-batch strategies and the training data can be reduced according to the fraction of low frequencies being kept: an exact $k$-truncated SVD has a complexity about $\mathcal{O}(kN^2)$. We note that fast $k$-truncated $\epsilon$-approximate SVD algrotithms for sparse matrix exist \citep{allen2016lazysvd}: if $\rho$ is the number of non-zero coefficients of $\tilde A$, the complexity can be about $\mathcal{O}(\frac{k\rho}{\epsilon}+\frac{k^2N}{\epsilon})$.
% \subsection{Taylor Expansion}
% \label{sec:taylorexp}

% We recall that $\tilde A_\eta = \frac{1}{1+\eta}\big(\eta I_N + \tilde A \big)$.
% Thus an eigenvector $u_k$ of $\tilde A$ corresponding to the eigenvalue $\lambda_k$ is an eigenvector of $\tilde A_\eta$ corresponding to the eigenvalue $\frac{\eta + \lambda_k}{1+\eta}$. Therefore, we have: 
% \begin{align*}
% \tilde A_\eta = \sum_{\lambda_k} \frac{\eta + \lambda_k}{1+\eta} u_ku_k^T\,,
% \end{align*}

% And thus,
% \begin{align*}
% \tilde A_\eta^n & =  \sum_{\lambda_k} \left(\frac{\eta + \lambda_k}{1+\eta}\right)^n u_ku_k^T\, \\
% \tilde A_\eta^n & =  \sum_{\lambda_k = 1} u_ku_k^T
% + \sum_{\lambda_k < 1} \left(\frac{\eta + \lambda_k}{1+\eta}\right)^n u_ku_k^T \\
% \tilde A_\eta^n & =  \sum_{\lambda_k=1}u_ku_k^T + \sum_{\lambda_k< 1}u_ku_k^T - \frac{n(1-\lambda_k)}{\eta} u_ku_k^T +\mathcal O(\frac{1}{\eta^2})
% \end{align*}
\end{document}